# Trustless parallel local search for effective distributed algorithm discovery

Zvezdin Besarabov
University College London
London, United Kingdom
zcabzbe@ucl.ac.uk

Todor Kolev
Comrade Cooperative
Sofia, Bulgaria
t.kolev@comrade.coop

## ABSTRACT

Metaheuristic search strategies have proven their effectiveness against man-made solutions in various contexts. They are generally effective in local search area exploitation, and their overall performance is largely impacted by the balance between exploration and exploitation.

Recent developments in parallel local search explore methods to take advantage of the efficient local exploitation of searches and reach impressive results. This however restricts the scaling potential to nodes within a private, trusted computer cluster.

In this research we propose a novel blockchain protocol that allows parallel local search to scale to untrusted and anonymous computational nodes. The protocol introduces publicly verifiable performance evaluation of the local optima reported by each node, creating a competitive environment between the local searches. That is strengthened with economical stimuli for producing good solutions, that provide coordination between the nodes, as every node tries to explore different sections of the search space to beat their competition.

## 1 INTRODUCTION

Metaheuristic search strategies allow for an automated methodology to effectively search for approximative solutions in a specific problem search space. Recent search strategies [1] [2] [3] in particular have reached impressive results in finding solutions beating the best man-made algorithms at the time.

Most well-known metaheuristic algorithms spend on average 90% of their execution time exploiting the search space [4]. Balancing exploration and exploitation is a critical question in their design. Most new generation metaheuristics approach this balance by first exploring and then switching the focus to local exploitation upon specific conditions [5]. This timing is crucial and optimizing it can yield significantly better results, the findings conclude.

This is primarily a problem for sequential search approaches however. Using parallelism introduces the hypothesis that we can take advantage of the efficient local search area exploitation of sequential searches, by executing many of these searches in different areas of the search space concurrently. Similar research has shown impressive results [6], especially when we are further optimizing how, when, and where we are starting and ending the local search processes. Increasing the number of processes also seems to correlate with increments in performance.

Scaling similar parallel local search systems requires a trusted network of computers, where all machines are truthfully executing their local search process and are reporting the correct findings. This restricts the scaling potential of similar approaches to centralized resource pools, such as data centers.

With the advent of Blockchain technologies [7] we have received the ability to create publicly trusted and verifiable state machines for arbitrary tasks [8], for example in finance, governance, and supply chains. This could also be applied to bring trust in a network of untrusted local search processes.

In this paper we hypothesize that we can utilize these technologies to design a parallel local search framework where the individual computational nodes are untrusted, arbitrary, and pseudo anonymous. This can practically allow for higher scalability as many more parties have the ability to contribute to the search. In addition, the use of Blockchain alleviates the need to have a centralized controller which orchestrates the local search processes. These rules can be enforced directly through the Blockchain state machine. If such a system is designed, this can mean that an arbitrary set of computing nodes can come together and find effective solutions to a given common problem without requiring trust or a governor. If we also add economical incentives for finding solutions, such a framework could become a self-sustaining, self-initiating environment where parties come to work on new solutions to problems organically.

More precisely, we are looking to devise a protocol for a Blockchain network where independent parties are incentivised to run a local search algorithm in a different point in a common search space and compete with the others in finding better solutions. The participants compete by exchanging proofs about the performance of their local optima, and they use this to decide whether to continue further optimizing in their local area or restart their search algorithm in a different one. This mechanism implies that it is in the competitors' best interest to explore new areas of the search space in order to gain a competitive advantage, and they have to collaborate on a feedback level to do so. From this point of view, this system represents an embarrassingly parallel local search system.

### 1.1 Related work

The current hypothesis has been motivated from our previous work with developing predictive strategies for cryptocurrency markets [9]. While applying a novel Neural Architecture Search approach [3] for the problem, we often reached some of the efficiency and scalability problems of similar sequential approaches. Systems for parallel hyperparameter tuning and inference have been previously explored before with largely positive results [10] [11]. For example, a "hopper" based neural network parallelization trainer scored a tenfold reduction in time compared to standard parallelism [12]. Recent developments in classical approaches such as Cooperative Co-Evolution [13] and Multi-population Differential Evolution [14] demonstrate how parallelism can be a natural next step in improving similar approximative searches. We have also seen the development of competitive parallel optimisation strategies that beat the most performing metaheuristics at the time [15], which shows interesting research potential in other competitive approaches.

All of the aforementioned approaches focus on improving distrubuted search in a privately owned network of computers, but there is recently growing demand for more open and shared ecosystems. Coin.AI [16] presents a novel Useful Poof-of-Work Blockchain consensus mechanism based on distributed neural network training, with possible applications in accelerating algorithmic search. Another interesting concept is Hydra [17], which presents a crowdsourced and fault tolerant way to scale neural model training and data collection from an arbitrary set of untrusted devices. In a similar manner OpenMined [18] allows models to be trained on a private, distributed dataset, and SingularityNet [19] creates a decentralised marketplace for algorithms as a service. Perhaps the closest match to a parallel local search is the platform Kaggle [20], which runs centralised and public competitions for the creation of better algorithmic solutions.

These approaches are either accelerating search strategies in a private set of computers, or scale openly on untrusted nodes to improve algorithm tuning without direct focus on an automated search process. The need for a solution which focuses on algorithmic search, scales to untrusted participants, and takes advantage of a competitive and economically stimulated environment led us to our current hypothesis.

## 1.2 Section overview

Section 3 introduces the novel blockchain protocol for a scalable parallel local search system by creating an improved environment for the execution of search strategies. This is defined more formally in Section 4, where we describe this functionality as rules of a blockchain state machine for the Tendermint framework, and Section 5 performs security analysis on the limitations of the protocol.

## 2 BACKGROUND
### 2.1 Blockchain

Satoshi Nakamoto's introduction of Bitcoin in November 2008 [7] has often been hailed as a radical development in money and currency, but the application of Blockchain as a tool for distributed consensus reaches far beyond. Later came Ethereum [8] with the ability to run custom blockchain applications within it. On a high level, a blockchain is a set of nodes that store and execute the same state machine with the same data. When someone wants to interact with it, they send a state machine transition to all nodes. This is checked for validity by every node and executed against the state machine. A few recent examples of its applications are within the areas of IoT security [21], IoT monetization [22], 5G communication [23], data source verification to combat Deepfake videos [24], and various Machine Learning approaches [25].

### 2.2 Tendermint

Tendermint [26] is a framework for building a blockchain network. Given an arbitrary state machine, it provides all other functionality necessary for running a blockchain - P2P networking, transaction creation, signing, and propagation, a consensus engine for block creation, propagation, and verification, as well as deterministic synchronisation of the machine state following block creation. The consensus is Byzantine fault tolerant as it can tolerate the arbitrary failure of up to $\frac{1}{3}$ of the participating peers.

A network based on Tendermint works as follows. First, nodes submit transactions that reach every other node in the P2P network and enter the nodes' *mempool*. Every node has a specific amount of "consensus power", and the rules for acquiring it are defined by the developer in the state machine. Upon a configurable event, a deterministic decision function is executed that selects one node on the base of its consensus power. This node selects transactions from the *mempool*, creates a block, and submits it to the network. Blocks are created on a specific blocktime schedule. The block has to be signed by nodes with at least $\frac{2}{3}$ of the network's total consensus power in order for it to be accepted on the blockchain. If this step fails within a timeout, a new block creator is selected. Nodes with non-zero consensus power are called blockchain validators.

Any functionalities outside of these, for example issuing and transacting a crypto asset, or "staking" assets to gain consensus power, can be freely implemented as rules within the state machine itself.

## 3 PARALLEL LOCAL SEARCH IN A BLOCKCHAIN

The idea of the protocol, which we will call "ScyNet", is to create an open environment where financially incentivised computing nodes can execute local search algorithms and compete with each other in finding optimal solutions to a specific problem. This happens with minimal coordination between the participants - everyone knows the bounds of the search space and the problem definition, but the only information exchanged throughout the search are proofs of the performance of the local optimum of the participants. This feature brings the self-governing aspect that omits a centralized controller to guide the local searches. If one node is consistently underperforming compared to its rivals, then it is stuck in a bad local optimum and should restart its local search in a different, possibly random point in the search space. This method of organization is more organic, spontaneous, potentially less computationally efficient, but allows us to scale to an unbounded number of untrusted search processes, which inevitably increases the probability of finding the global optimum.

The general concept of ScyNet is visualized in Figure 1, with further clarifications in the following sections.

### 3.1 Main entities and roles

A specific implementation of ScyNet for a chosen problem (for example, stock market predictions) represents a single blockchain network and is called a **Domain**. Every domain defines an unique utility token that is used to form consensus and incentivise participants. Three types of nodes, members of the network, exist - data nodes, searchers, and clients. Data nodes interface the real world by selling data related to the domain problem. Searchers can use that data to execute a local search algorithm that creates algorithmic **agents** as optimal solutions to the problem. With every announced agent, searchers submit cryptographic proofs of the agents' performance. Then, clients can explore these proofs and can contact specific searchers to request access to the produced solutions, for example by paying a subscription. This provides the crucial economical incentive for searchers, and therefore data nodes, to invest resources in this problem in the first place.

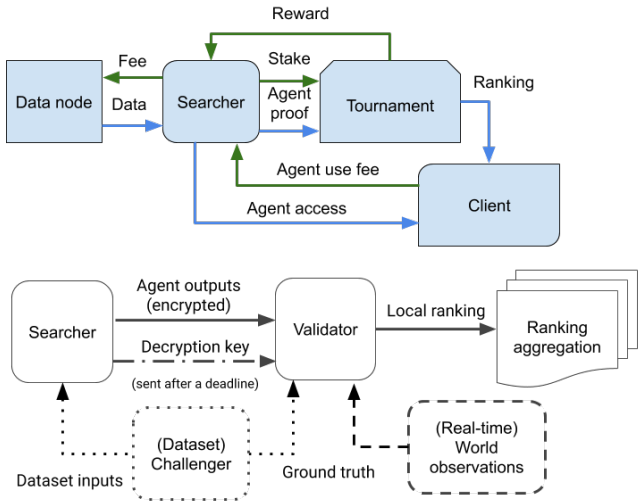

Figure 1: Top: A high-level overview of the main participants, interactions, and lifecycle of ScyNet.
Bottom: How Agent performance is verified during a tournament. Depending on the domain, this includes either a challenger or an external source of real-time ground truth.

### 3.2 Tournament validation

Tournaments are the way searchers exchange performance proofs to "verify" their agents, and hence compete. They are regularly scheduled and continuously running - a tournament has a start date, which is right after the previous one ends. Searchers have the right to submit one or more of their agents for verification by paying a submission fee before the start of the next tournament.

Tournaments cryptographically prove an agent's performance when searchers submit their agents' outputs, or "signals", given specific input data or conditions. This depends on the type of algorithmic problem being solved in the domain. The first type represents real-time predictions, such as stock market trading or weather forecasts. The problem definition includes a consistent schedule, known as real-time ticks, when such predicions should be made, and everyone knows how to get the current ground truth of the prediction target. During a tournament, the competing searchers are required to provide real-time predictions from their agents for every tick during the tournament. When the ground truth becomes available, a subset of the network nodes, validators, evaluate the accuracy of the predictions to form consensus on a ranking of the participating agents.

The other type of problems are dataset input-output problems, such as self-driving algorithms or speech recognition. In this case, the blockchain state machine pseudorandomly selects multiple nodes that are labeled as **challengers**. Every such node is responsible to provide a dataset with which to challenge the competing agents to resolve it. Depending on the problem, this can be either algorithmically generated, or retrieved from a reliable source. The searchers inference the dataset and publish the outputs of their agents. After tournament closure, the validators compare the agent outputs to the challenger ground truth and forms consensus on an agent ranking.

All submitted agent outputs are first encrypted to avoid copying. The submission fees collected before a tournament comprise the tournament award, which is used to award the top performing agents and the selected challengers, if applicable.

At no single point do the searchers reveal the mechanisms behind their agents, allowing for a significant flexibility in how these agents are created.

### 3.3 Agent utilization

Since searchers retain ownership of their agents, they can monetize them in arbitrary ways, including offering them as a service via subscription or per-use fees, or selling the underlying algorithm altogether.

In a real-time predictive domain, it may also be possible for a searcher to subscribe to the prediction stream of an agent of another searcher, and use this as input data to its own agents, hence possibly making a more informed prediction. This type of aggregation has also been used previously to deliver explainable AI [27].

## 4 FORMAL SCYNET PROTOCOL DEFINITION

This section extends on Section 3 by clarifying how the presented features are to be implemented and enforced. The ScyNet blockchain protocol is designed as a Tendermint state machine and a set of transactions to interact with it. Here we will describe the logical rules of this state machine in relation to which transactions are sent and when, as well as the validity rules for these transactions.

The blockchain protocol defines a trustless way in which a group of nodes that internally execute local searches can agree on a ranking between their local optima. We are assuming that software is present on every node, not part of the blockchain consensus, that defines and supports the following:

- The data formats used for datasets and agent signals;
- The error function for performance evaluation;
- Software to run, manage, and reinstantiate the local search based on feedback from blockchain tournaments;
- For dataset domains, a method for generating or gathering a new dataset on demand if the node is selected as a challenger;
- Optionally, access to correctly formatted data that can be sold on the blockchain if this node becomes a data node.

## 4.1 Underlying blockchain

On a lower level, a ScyNet domain builds an independent blockchain network and a blockchain structure. The network behaves similarly to Bitcoin [7], where any external node can connect with other peers, synchronize the blockchain, interact with the network by signing transactions, as well as participate in the consensus by verifying transactions and blocks.

*4.1.1 Token.* Every domain network creates a non-mintable token of a fixed supply. This token is used in all network payments to incentivise the consensus and commitments of the three types of nodes - data nodes, searchers, and clients.

*4.1.2 Consensus.* By using Tendermint, most aspects of a consensus mechanism are already provided - block creator selection, verification, propagation, as well as Byzantine Fault Tolerance to various attacks. What has to be specified by us is how network participants gain network consensus power. While this is not a focus in our protocol as there are many good implementations based on the use case, we are recommending the use of a variant of Proof-of-Stake [28] called "Coin Age based selection", as used in PeerCoin [29]. The consensus power is a function of the balance a node has "staked" and the time since it did that. However any of the more prominent PoS examples can be implemented here, including more recently Snow White [30].

## 4.2 Domain configuration

The definition of a domain network includes setting a few specific parameters. This defines the tournament schedule and validation parameters, which are part of the problem definition:

- problemType ("real-time" or "dataset") - Type of algorithmic problem;
- tournamentStartFrequency (Integer, milliseconds) - The interval between the start of two consecutive tournaments;
- challengerSubmissionTimeout (Integer, milliseconds) - Timeout for the the selected network challengers to submit a dataset;
- datasetSignalKeyTimeout (Integer, milliseconds) - Timeout for challengers and searchers to send dataset signal decryption keys;
- rankingTimeout (Integer, milliseconds) - Timeout after the tournament where validators must submit tournament results;
- realTimeFrequency (Integer, milliseconds) - For real-time problems, the interval between two real-time ticks;
- minAgentChallengers (Integer) - Minimum number of nodes that have to independently challenge a submitted agent;
- minAgentChallengerVotingPower (Percentage) - Minimum share of the network voting power that has to challenge a submitted agent;
- minAgentSubmitStake (Domain tokens) - Minimum stake for a searcher to submit an agent for validation in a tournament;
- minPricePublishStake (Domain tokens) - Minimum stake for a searcher or data node to advertise their agent/data on the chain;
- rentFee (Domain tokens) - Fee withdrawn when a client rents an agent.

## 4.3 Timing constraints

The start timestamp of all tournaments is defined as UNIX Epoch + k*$tournamentStartFrequency$ for all $k > 0$. Also, if the domain is real-time, the generalized timestamp of all real-time ticks is UNIX Epoch + m*$realTimeFrequency$. All other timing constraints during the ScyNet livecycle are relative to the start of a specific tournament and are defined as follows:

- tournament end = the tournament start timestamp of the next tournament;
- challenger submission deadline = tournament start + $challengerSubmissionTimeout$;
- dataset key deadline = tournament end + $datasetSignalKeyTimeout$;
- ranking deadline = tournament end + $rankingTimeout$;

When we describe that a certain event happens at/after/before a given timestamp, this event should have happened at/after/before a node receives a block with the given timestamp.

## 4.4 Role protocol

This section defines the legal way for a node with a particular role to interact with the blockchain state machine, indirectly defining the machine itself. References will be made to both the domain configuration (Section 4.2) and the different network transactions (Section 4.5). Network participants may execute multiple roles at the same time (for example, to be a searcher and a data node).

*4.4.1 Data node protocol.* In order for a node to become a data node, provide a data agent (real-time stream or a specific dataset), and advertise it on the blockchain, it must:

(1) Submit publish_data_price transaction (Section 4.5.8). This stakes (or locks) a voluntary amount of tokens $\geq minPricePublishStake$, demonstrating confidence in the data agent;
(2) Listen for rent transactions (4.5.9) that are directed to the node;
(3) Communicate with the sender off-chain and privately give access to the requested quantity or duration of data.
(4) When the node wants to stop providing a specific data agent, it sends publish_data_price transaction (4.5.8) with a stake of 0. This will declare the agent as unavailable and return the node's stake.

*4.4.2 Searcher protocol.* In order to verify and sell an agent on the network, a searcher should:

(1) Send submit_agent transaction (4.5.1). This stakes a voluntary amount of tokens $\geq minAgentSubmitStake$;
(2) On the next tournament start, the blockchain state is updated so that agents with within the highest $tournamentCutoff$ % stake are allowed to participate in the tournament.
(3) If the searcher's agent is participating and domain is real-time:
   (a) Wait until right before every specific real-time tick during the tournament;

(b) Generate an AES-256 key, generate an agent signal, encrypt the signal, and send submit_signal transaction (4.5.3) before the real-time tick;
(c) After the current real-time tick and before the next one, reveal the decryption key with publish_signal_decryption_key transaction (4.5.5);
(d) Repeat from 3a for every real-time tick until tournament end.

(4) If the searcher's agent is participating and domain is dataset:
  (a) Wait tournament start, after which a distributed pseudorandom algorithm selects the tournament challengers. Listen for publish_dataset transactions (4.5.2) from the challengers until the challenger submission deadline;
  (b) Download the datasets, generate all agent signals, generate an AES-256 key, encrypt the signals, and send submit_signal transaction (4.5.3) before tournament end;
  (c) Wait for tournament end and reveal the decryption key with publish_signal_decryption_key transaction (4.5.5) before the dataset key deadline.

(5) After successful validation and if the searcher decides, they may submit publish_agent_price transaction 4.5.7 to advertise the monetization of their agent, also placing a voluntary stake;

(6) Listen for rent transactions (4.5.9) that are directed to the searcher;

(7) Communicate with the sender off-chain and privately give access to the requested agent use.

(8) When the searcher wants to stop providing a data agent, it sends publish_agent_price transaction (4.5.7) with a stake of 0. This will declare the agent as unavailable and return the searcher's stake.

4.4.3 **Challenger protocol (dataset domains).** Challenger selection is similar to the way Tendermint selects a pseudorandom block creator given a set of nodes weighted by consensus power, except that here multiple nodes are selected and labeled as challengers.

Challengers are selected right after tournament start as follows:

(1) Retrieve a list of current network nodes, sorted by their power;
(2) Hash the previous block's header and the consensus signatures in the current block;
(3) Use that hash as a seed in a deterministic RNG algorithm and generate a pseudorandom integer in the interval $\in (1, totalValidatorPower)$;
(4) Start iterating the node list, summing their power until the sum surpasses the selected integer;
(5) Mark the node where we stopped iterating as a challenger if they are not a searcher participating in the tournament;
(6) Loop from step 3 until:
  (a) There are at least $minAgentChallengers$ challengers selected;
  (b) The total consensus power of selected challengers is at least $minAgentChallengerVotingPower$ %;
  (c) There is not a challenger that comprises more than 10% of the total challenger consensus power;

Once selected, every challenger must:

(1) Run a domain-specific algorithm to generate a validation dataset;
(2) Generate an AES-256 key, encrypt the dataset outputs, and submit the dataset and its encrypted outputs through the publish_dataset transaction(4.5.2) by the challenger submission deadline;
(3) Wait for tournament end and submit dataset outputs decryption key via publish_dataset_decryption_key (4.5.4) before the dataset key deadline.

4.4.4 **Common protocol for blockchain validators.** Other than building the underlying consensus, blockchain validators also judge the performance of the agents in a tournament:

(1) If domain is real-time:
  (a) Listen for submit_signal transactions (4.5.3) from the searchers;
  (b) Listen for publish_signal_decryption_key transactions (4.5.5);
  (c) Wait for tournament end and decrypt all received signals;
  (d) Make a local ranking of agents based on their prediction accuracy.
(2) If domain is dataset:
  (a) Listen for publish_dataset transactions (4.5.2) from the challengers;
  (b) Listen for submit_signal transactions (4.5.3) from the searchers;
  (c) Listen for publish_dataset_decryption_key (4.5.4) from the other challengers;
  (d) Listen for publish_signal_decryption_key transactions (4.5.5) from the searchers;
  (e) Decrypt all received signals and make a local ranking of the agents based on their median score among all challenger datasets;
(3) Submit the local ranking in a publish_tournament_ranking (4.5.6) transaction. It must be included in a block with a timestamp before the ranking deadline.

4.4.5 **Common protocol for every node.** Other than syncing with the blockchain, every node in the network also ensures other nodes are following the requirements of their roles. If a node misbehaves, they or their agents are marked as disqualified on the blockchain. Disqualifications affect the node's eligibility to receive the tournament award, as well as their ability to continue sending specific transactions (Clarification in Section 4.5). Disqualifications last only for the specific tournament.

(1) Mark any competing in a tournament agent as disqualified if any of the following is true:
  (a) In a real-time domain, the searcher did not send a valid submit_signal (4.5.3) or publish_signal_decryption_key (4.5.5) transaction for every real-time tick;
  (b) In a dataset domain, the searcher did not send a valid submit_signal (4.5.3) by the tournament end or

publish_signal_decryption_key (4.5.5) by the dataset key deadline.
(2) In a dataset domain, mark any challenger as disqualified if any of the following is true:
   (a) The node did not send a valid publish_dataset (4.5.2) by the challenger submission deadline or publish_dataset_decryption_key (4.5.4) by the dataset key deadline.
(3) After the proposer deadline for a tournament, mark every validator that did not send a valid publish_tournament_ranking (4.5.6) for the specific tournament as disqualified.
(4) For the duration of the tournament, let $bl$ denote the amount of blocks created by a specific validator and $txs$ - the number of all transactions included in them. For every network validator, calculate $bl + txs$ after tournament end. This called block creator points and is later used in the tournament reward;

4.4.6 Tournament reward. After the ranking deadline, all agent submission fees that comprise the tournament reward are used to reward the participants in the tournament. This includes the winning agents, participating challengers, and validators that created blocks during the tournament. The funds are allocated as follows:

- 10% of the funds are split among all non-disqualified validators that created blocks during the tournament proportional to their block creator points;
- If the domain is dataset, another 10% of the total reward is split among the participating non-disqualified challengers in proportion to their consensus power;
- 5% of the total reward is sent to a predefined address, responsible for maintaining and further expanding the ScyNet protocol and implementation;
- The rest of the funds are allocated among the top 50% non-disqualified agents in ratio of their scores.
- If all agents have been disqualified, the remains of the reward are transferred to the next tournament reward.

## 4.5 Transaction types

All interaction with the blockchain state machine is executed through transactions. In the following subsections we very briefly summarize the types of transactions used in this protocol. A more complete description and validity constraints is available in the following document [31].

4.5.1 submit_agent(*UUID*, *stake*). Sent by a searcher node to notify that it has an agent that it wants to verify in the next tournament.

4.5.2 publish_dataset(*inputsURL*, *inputsHash*, *encryptedSignalsURL*, *signalsHash*). Used by tournament challengers to publish the inputs of their personal dataset.

4.5.3 submit_signal(*agentUUID*, *encryptedSignal*). Sent by a searcher node to submit an AES-256 *encryptedSignal* from a specific agent (*agentUUID*).

4.5.4 publish_dataset_decryption_key(*key*). Sent by tournament challengers after the end of a tournament to reveal the decryption key behind publish_dataset.

4.5.5 publish_signal_decryption_key(*agentUUID*, *key*). Sent by a searcher a specific amount of time after signing submit_signal to reveal the AES-256 key by which the original signal was encrypted.

4.5.6 publish_tournament_ranking(*ranking*). Submits the local tournament ranking of one validator.

4.5.7 publish_agent_price(*agentUUID*, *scheme*, *price*, *stake*). Advertises purchasing rules for a previously verified agent.

4.5.8 publish_data_price(*dataUUID*, *dataParams*, *scheme*, *price*, *stake*). Advertises purchasing rules for a dataset.

4.5.9 rent(*UUID*, *quantity*). Purchases access to a previously advertised agent or data.

## 5 ATTACK RESILIENCE

In this section, we describe how the specifics of the ScyNet transaction protocol build resilience against various types of failures.

## 5.1 Underlying blockchain security

The Tendermint framework allows for the synchronous processing of transactions that propagate through the network in a deterministic manner. There is no disparity in the system, meaning that replay attacks are not possible. Because of the voting mechanism for consensus, if one party obtains $\geq \frac{1}{3}$ of the network's consensus power, it can stop block verification. The system can be arbitrarily modified with $\geq \frac{2}{3}$ of the consensus power.

Additionally, the transaction signing makes impersonation not possible without access to a node's private key.

## 5.2 Transaction spam

In the detailed description of the transactions [31], every transaction type that does not require a fee has constraints on its usage for a time period. Sending a transaction outside of these limits will invalidate it. Tendermint nodes do not propagate invalid transactions they received, limiting the extent of similar attacks. Transactions with fees have monetary incentives against this behavior. This builds resilience against Denial-of-Service transaction attacks.

## 5.3 Agent signal copying

A searcher may be tempted to copy a rival's agent signals during a tournament and submit them as their own. However, all agent signals are submitted encrypted. The decryption key is revealed after the deadline for submission of a specific signal, protecting against copying.

A searcher may instead copy all submitted ciphertexts of a rival and then copy the decryption key. However, when submitting a signal, the searcher is required to encrypt both the signal and their public key, signing the package with their private key.

## 5.4 Service failure

If a client has paid a searcher or a data node to utilize its agent or data, the providing node can theoretically deny access to the service, as it is not enforced by the blockchain consensus. However, this will damage the node's reputation and therefore its ability to monetize its past and future agents, which likely diminishes the reason to create them in the first place. It has to be noted that for this mechanism to work, the nodes that sell data and agents should not be anonymous and they should establish real-world reputation.

## 5.5 Agent submission failure

A searcher may submit an agent for participation in a tournament and then fail to send a required timed agent signal or decryption key. In this case, the they only disadvantage themselves and will be disqualified.

## 5.6 Challenger failure

In dataset domains, failing to send a valid, readable, dataset or an output decryption key by the deadlines will disqualify that challenger.

Submitting a valid dataset with incorrect outputs is possible, as the network does not distinguish between an inaccurate agent and bad testing data. The challenger can also secretly provide the unencrypted correct outputs to a competing searcher, giving them an unfair advantage. However, because the agent's performance is its median from all challenger datasets, the performance measurement is resilient up to the arbitrary failure of half of all challengers. Section 4.4.3 also defines restrictions to ensure that the amount of challengers is always sufficiently large.

# 6 LIMITATIONS

Overall while very scalable, because of the limited communication, this trustless parallel local search approach is likely to lose computational efficiency the more local searches there are. This is because there is no cheap and direct way in which a new network member can ensure that they are exploring a brand new section in the search area. This should become less of a problem in more complicated search areas, such as with Neural Architecture Search [1] [2] [3], where starting from a random search point is more likely to be unique anyways.

Currently the protocol is designed for searches for data-based approximative solutions, which are either making real-time predictions from new data, or inferencing a specific dataset. Due to network restrictions very fast-paced or low-latency predictive problems are infeasible. This is also not suitable for dataset problems where there is no practical way in which a node can create or source a piece of testing dataset which is not already publicly known. Sensitive or private data is also not directly applicable, but this can be mitigated with homomorphic encryption and federated learning approaches as in [18].

# 7 FURTHER WORK

While the protocol has been proposed, it must be more thoroughly analysed and simulated, especially to prove that the potential attacks in Sections 5.2, 5.4, 5.6 are economically infeasible. The protcol can also be expanded to support data-free problems, such as Reinforcement Learning solutions, by allowing challengers to run a RL environment instead of sourcing a dataset.

In longer-term, our plans are to use the presented blockchain protocol to implement a public blockchain network, which we will also call "ScyNet".

Even with a small number of initially participating nodes, ScyNet can be a meaningful source of algorithmic solutions that is entirely automated and autonomous. The only resource that this network requires to produce continuously improving solutions is computing power.

# 8 CONCLUSION

We presented an open competitive trustless environment of algorithm creators, which perform local search strategies and coordinate their performance results with the other participants. This in turn allows said participants to select their next search area more efficiently.

The biggest contribution of this research is the creation of the publicly verifiable secure evaluation function behind the blockchain tournaments. This removes the need of trusting third parties or other participants in the network and provides a fair and unbiased competitive environment. Coupled with the aforementioned mechanism for search space diversity, this trustless system can scale to a massive number of parallel local searches, further increasing the probability of finding the global optimum.

While this is applicable in various searches for approximative algorithms, we believe that the most interesting application is in the domain of Neural Architecture Search (NAS). Research in the area has emphasized on the great potential of NAS and its current limitations on scalability and search space diversity [1] [2] [3]. A good further hypothesis is to analyze if the performance of NAS can be improved through our approach.

Our future work will be towards providing a reference implementation of the presented blockchain protocol and establishing an operational network of computing nodes.